# Effect of Heavy-Ion Irradiation on Superconductivity in Single Crystals of $Ba_{0.6}K_{0.4}Fe_2As_2$ Pnictide Superconductor

N. W. Salovich[1], H. Kim[2], A. K. Ghosh[1,3], R.W. Giannetta[1], W. Kwok[4], U. Welp[4], B. Shen[4], S. Zhu[4], H.-H. Wen[5,6], M. A. Tanatar[2], R. Prozorov[2]

[1] *Loomis Laboratory of Physics, University of Illinois at Urbana-Champaign, Urbana, IL 61801, USA*

[2] *Ames Laboratory, Department of Physics and Astronomy, Iowa State University, Ames, Iowa 50011, USA*

[3] *Department of Physics, Jadavpur University, Kolkata 700032, India*

[4] *Argonne National Laboratory, Argonne, IL 60439, USA*

[5] *Center for Superconducting Physics and Materials, National Lab for Solid State Microstructures*

[6] *Department of Physics, Nanjing University, Nanjing 210093, China*

## Abstract

The London penetration depth was measured in optimally doped $Ba_{0.6}K_{0.4}Fe_2As_2$ crystals, with and without columnar defects produced by 1.4 GeV $^{208}Pb$ irradiation. The low temperature behavior of unirradiated samples was consistent with a fully gapped superconducting state with a minimum energy gap $\Delta_{min}/k_B T_C \approx 1$. Similar gap values were observed for irradiation levels corresponding to mean column-column separations of 32 nm and 22 nm. At very high irradiation levels (column-column separation of 10 nm) a $T^2$ power law was observed below $T_C/3$, most likely due to elevated scattering. Neither the location nor the sharpness of the superconducting transition was affected by irradiation. The data provides evidence for an $s_{+-}$ pairing state.

A multigap, sign-changing order parameter termed the $s_{+-}$ phase [1,2] is a leading candidate for the pairing state in iron-based superconductors. Although there is considerable evidence for multiple energy gaps[3], direct proof of a sign change remains a challenge. Several authors [4,5] have predicted that impurities should change the temperature dependence of various quantities in a manner that is sensitive to the relative sign. For example, the London penetration depth in an $s_{+-}$ state is expected to evolve from an exponentially activated (BCS-like) dependence to a $T^2$ power law with increasing impurity concentration [4,5]. The effect of impurities on $T_C$ is unsettled. Originally, non-magnetic interband scattering was predicted to rapidly suppress $T_C$ [6] but this conclusion was later revised [7,8]. Isolating the role of impurities is difficult since they may also change the carrier concentration and in turn change the pairing state. Indeed, recent thermal conductivity measurements in $Ba_{1-x}K_xFe_2As_2$ show a transition from a nodeless state to nodal d-wave pairing upon doping towards pure $KFe_2As_2$ [9,10]. Columnar defects produced by heavy ion irradiation offer an alternative way to test the effect of impurity scattering. Columns do not ostensibly change the carrier concentration or add magnetic scattering centers and their density may be reliably controlled. Since columns are also effective vortex pinning centers [11,12] their effect on the superconducting properties is important to understand.

In this letter we report penetration depth measurements on single crystals of $Ba_{0.6}K_{0.4}Fe_2As_2$, a hole-doped iron-based superconductor [13]. By studying both pristine and irradiated samples taken from the same crystal we isolate the effect of columnar defects. This approach was first carried out on the related electron–doped superconductors $Ba(Fe_{1-x}Co_x)_2As_2$ and $Ba(Fe_{1-x}Ni_x)_2As_2$ where changes in $T_C$ and penetration depth were in agreement with an $s_{+-}$ pairing model [14]. For $Ba_{0.6}K_{0.4}Fe_2As_2$, we find that heavy ion irradiation does not change $T_C$, even with a column to column separation of only a few coherence lengths. However, very dense columnar defects do cause the penetration depth to acquire a $T^2$ power law dependence, which is strong evidence for an $s_{+-}$ state. The power law is critical since the penetration depth remains exponential at low temperatures even in the dirty limit of a conventional isotropic s-wave superconductor [3].

Measurements were performed on two groups of (nominally) optimally doped single crystal $Ba_{0.6}K_{0.4}Fe_2As_2$ [15]. Irradiation with 1.4 GeV $^{208}Pb$ ions was performed at the Argonne Tandem Linear Accelerator using a gold foil to disperse the beam via Rutherford scattering to ensure a uniform beam spot over the sample. The beam current was held below 500 pA to avoid sample heating. Heavy ions formed tracks along the c-axis with an average stopping distance of 60-70 $\mu m$, larger than the thickness of the crystals. The irradiation level is specified by fluence, mean column-column separation $r$ and the matching field (in Tesla) defined as $B_\phi = \phi_0/r^2$ where $\phi_0$ is the flux quantum. For the first sample group (measured at Ames Laboratory) a single crystal with $T_C = 39$ K was cut into several smaller segments. One segment was left unirradiated ($B_\phi = 0$) while the other two segments were given irradiation doses of $B_\phi = 2$ T ($r = 32$ nm, fluence = 9.6 x $10^{10}$ ions/cm$^2$-sec) and $B_\phi = 4$ T ($r = 22$ nm, fluence = 1.9 x $10^{11}$ ions/cm$^2$-sec). The samples measured at UIUC were taken from a single crystal with $T_C = 36.8$ K. One segment was left unirradiated while the second was irradiated to $B_\phi = 21$ T ($r = 10$ nm, fluence = $10^{12}$ ions/cm$^2$-sec).

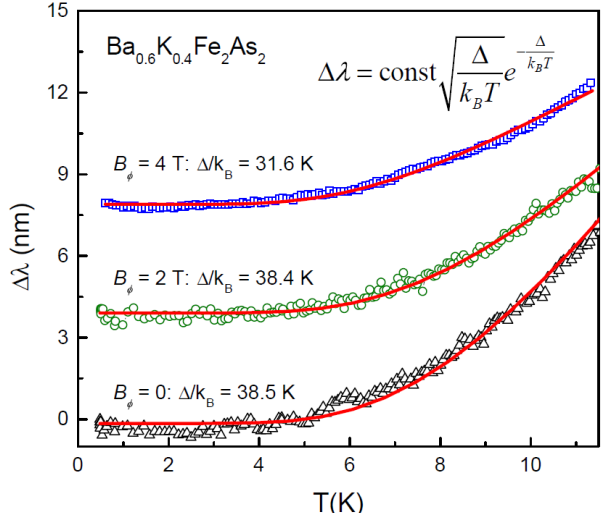

Fig. 1. Change in penetration depth for 3 samples from first group (Tc = 39 K) for columnar defect densities of $B_\phi = 0$ T, 2 T, 4 T. Fits to BCS temperature dependence are shown as solid curves.

Changes in the penetration depth with temperature were measured with a tunnel diode resonator described in several previous publications [16-18]. Two separate but functionally equivalent oscillators were used, each with base temperature of 400 mK. Upon *in situ* insertion of the sample into the probe coil, the shift in oscillator frequency $\Delta f$ is given by,

$$\Delta f = \frac{gV_S}{1-N}\left(1 - \frac{\lambda}{R}\tanh\frac{R}{\lambda}\right) \quad (1)$$

$R$ is an effective sample dimension, $N$ is the sample demagnetizing factor, $g$ is a calibration constant and $\lambda$ is the penetration depth for currents flowing perpendicular to the c-axis [17,18]. Figure 1 shows the penetration depth in the low temperature region ($T/T_C < 0.32$) for the first sample group ($T_C = 39$ K). Over the temperature range shown the data for each irradiation level was fit to a BCS-like form, $\Delta\lambda \sim \sqrt{\Delta/T}\, e^{-\Delta/T}$. The gap values were $\Delta_{min}/k_B T_C = 0.99$ ($B_\phi = 0$), 0.99 ($B_\phi = 2$ T) and 0.81 ($B_\phi = 4$ T). In each case the BCS expression provided a superior fit to a power law. These gap values are in good agreement with STM [15] and ellipsometry [19] measurements that report

$\Delta_{min}/k_B T_C = 1.1$. Figure 2 shows similar data for the second sample group ($T_C = 36.8\,K$). Data for the unirradiated sample was best fit to a BCS form with $\Delta_{min}/k_B T_C = 0.97$. These minimum gap values are all well below the weak-coupling BCS value of $\Delta_{min}/k_B T_C = 1.76$, implying at least two distinct energy gaps and consistent with many experiments in the iron-based superconductors [3].

The lower panel of Fig. 2 shows data for the heavily irradiated ($B_\phi = 21$ T) sample. In this case a $T^2$ power law provided a clearly superior fit to the BCS form. There was no evidence of a low temperature upturn that can arise from magnetic impurities [18]. Therefore scattering from the columnar defects should be regarded as nonmagnetic. Figure 3 shows data in the vicinity of $T_C$. For the first group (lower panel) the midpoint transition temperature of $T_C = 39\,K$ was unaffected by irradiation for all three matching fields. This data should be contrasted with a systematic suppression of $T_C$ for similar irradiation levels in Ba(Fe$_{1-x}$ T$_x$)$_2$As$_2$ (T = Co, Ni) [14]. For the second sample group (upper panel) the midpoint transition of $T_C = 36.8\,K$ was the same for both unirradiated and irradiated samples. The different $\Delta\lambda(T_C)$ values for various irradiation levels reflect differences in the effective dimension $R$ of the crystals.

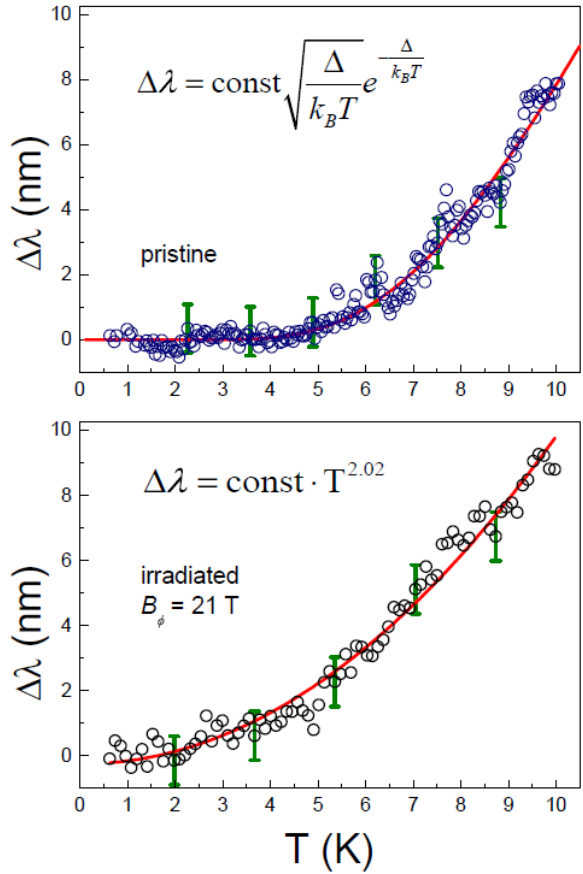

Fig. 2. Change in penetration depth for second sample group ($T_C = 36.8\,K$). (Upper) Data for unirradiated sample with BCS-like fit. (Lower) Sample with $B_\phi = 21$ T matching field showing quadratic power law fit.

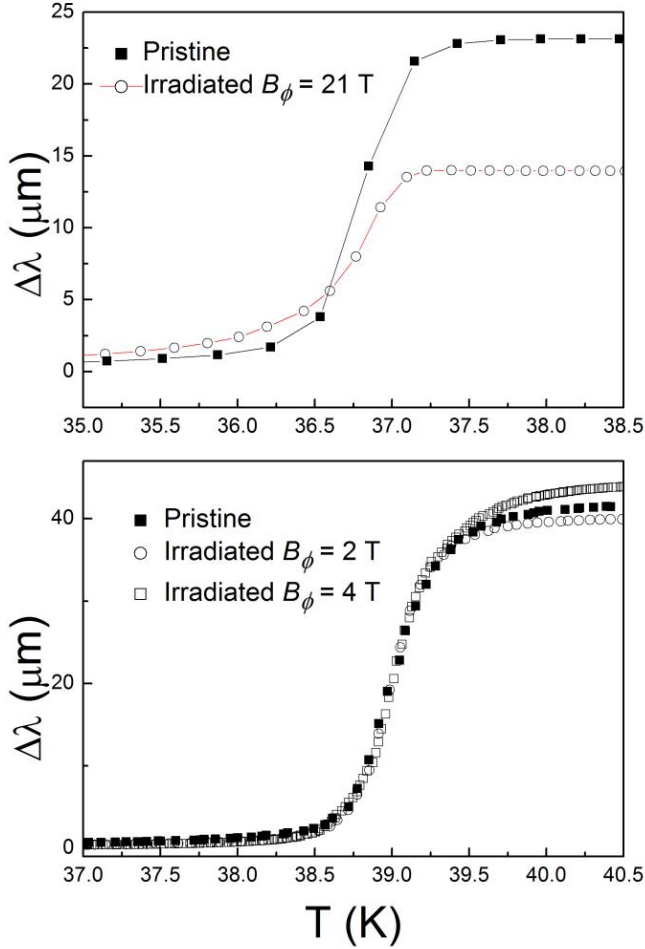

Fig. 3. Change in penetration depth near $T_C$ for both sample groups and different irradiation levels.

The highest irradiation level ($B_\phi = 21$ T) corresponds to an average column separation of 10 nm. TEM images show that the columns themselves have an average diameter of 3.7 nm. Were the tracks continuous, the superfluid would be confined to regions of order 6.3 nm or less, i.e. roughly 2 coherence lengths. However, the confinement is somewhat weaker since column tracks are actually discontinuous and a given cross section reveals a column density roughly 1/3 of what is expected from the dose matching field. At high doses there is also fairly high probability that an ion hits close to the track of a previous impact, leading to a nonuniform damage pattern with local strain fields.

The data in Figures 2 and 3 appear difficult to reconcile. The evolution from either $\Delta\lambda \sim \sqrt{\Delta/T}\, e^{-\Delta/T}$ or $\Delta\lambda \sim T^n\,(n>2)$ *toward* $\Delta\lambda \sim T^2$ with increased scattering is predicted to occur with an $s_{+-}$ order parameter for which interband scattering produces midgap states, but not for an $s_{++}$ pairing state [4,5]. However, interband scattering (unless it is in the pure unitary limit) should *also* suppress $T_C$. Experiments in ion-irradiated Ba(Fe$_{1-x}$T$_x$)$_2$As$_2$ (T = Co, Ni) show both effects and provide evidence for $s_{+-}$ pairing [14]. The much stronger $T_C$ suppression in Ba(Fe$_{1-x}$T$_x$)$_2$As$_2$ may be a result of its larger gap anisotropy, as revealed in both thermal conductivity and penetration depth measurements [20-22]. At least for Born-limit scattering, generalizations of the Abrikosov-Gor'kov [23] theory show that gap anisotropy, whether sign-changing or not, permits non-magnetic scatterers to reduce $T_C$ [24,25,26]. Therefore, if the defects produced by irradiation induce predominantly out-of-plane, small wave-vector (interband) scattering events, these will suppress $T_C$ in a material with an anisotropic gap, such as Ba(Fe$_{1-x}$T$_x$)$_2$As$_2$. The same scatterers will have little effect on Ba$_{0.6}$K$_{0.4}$Fe$_2$As$_2$ where the gap is much more isotropic. Increasing amounts of large wavevector (interband) scattering will eventually lead to midgap states and $\Delta\lambda \sim T^2$ in either superconductor if the pairing is $s_{+-}$ [27]. Hashimoto et. al. [28] performed penetration depth measurements on several unirradiated samples of Ba$_{1-x}$K$_x$Fe$_2$As$_2$ [28]. They also observed a crossover from exponential to $\Delta\lambda \sim T^2$ as the impurity scattering rate (as measured by microwave surface resistance) increased, again consistent with $s_{+-}$ pairing. In contrast to our data, they found that $T_C$ was suppressed from 32.7 K to 25 K with higher scattering rates. The difference suggests that either impurities produce stronger interband scattering than columnar defects or that the differences in $T_C$ in between samples reflected different effective doping levels.

A more quantitative understanding of scattering from columnar defects would clearly be desirable. However, some features may be independent of the details. For example, our data are consistent with recent calculations showing universal behavior in the $T_C$ suppression for $s_{+-}$ states [7,8]. Those authors identify two different $s_{+-}$ states, depending on the sign of the pairing interaction averaged over two bands. For an average repulsive interaction, $T_C$ suppression follows an Abrikosov-Gor'kov [23] scenario in which a generalized scattering rate of order $\hbar\Gamma \sim k_B T_{C0}$ drives $T_C$ to zero while maintaining $s_{+-}$ pairing. For a net attractive interaction, $T_C$ suppression is exceedingly weak and a universal function of scattering parameters. For attractive coupling, the state begins as fully gapped $s_{+-}$ ($\Delta\lambda \sim \sqrt{\Delta/T}\, e^{-\Delta/T}$). With increased scattering one gap crosses zero, leading to a concentration of midgap states ($\Delta\lambda \sim T^2$) as observed here. Further increase of the scattering leads to $s_{++}$ pairing and a return to $\Delta\lambda \sim \sqrt{\Delta/T}\, e^{-\Delta/T}$. The observation of this re-entrant behavior would require a defect density even larger than $B_\phi$ = 21 T.

The previous scenarios omitted mention of spin density wave (SDW) order, which may be an important variable. Both Ba(Fe$_{1-x}$Co$_x$)$_2$As$_2$ [29] and Ba$_{1-x}$K$_x$Fe$_2$As$_2$ [13,30,31] possess a SDW/SC coexistence region. Disorder generally suppresses the SDW transition and may simultaneously reduce or even enhance $T_C$ for an $s_{+-}$ state [32]. The detailed behavior of $T_C$ therefore depends on the level of doping as well as the character of the scattering, making this model difficult to assess. We stress that SDW/SC coexistence on a microscopic scale is itself a strong indication of $s_{+-}$ pairing [33].

It is useful to compare $Ba_{0.6}K_{0.4}Fe_2As_2$ with $MgB_2$, a phonon-mediated superconductor with weak anisotropy and two unequal energy gaps [34]. Chikumoto et. al.[35] irradiated $MgB_2$ samples with 5.8 GeV Pb ions and reported no change in $T_C$ for matching fields up to $B_\phi = 4$ T. This might suggest that $s_{++}$ pairing holds for $Ba_{0.6}K_{0.4}Fe_2As_2$ as it does in $MgB_2$. However, to our knowledge there is no report of an impurity-induced crossover from exponential to $\Delta\lambda \sim T^2$ in $MgB_2$. In addition, the lack of $T_C$ suppression in $MgB_2$ is likely a peculiarity of its weak interband scattering and not a general feature of $s_{++}$ pairing [36,37].

To summarize, $Ba_{0.6}K_{0.4}Fe_2As_2$ crystals show no change in $T_C$ for columnar defect densities up to $B_\phi = 21$ T matching field. This robust $T_C$ contrasts sharply with ion-irradiated $Ba(Fe_{1-x}Co_x)_2As_2$ suggesting that $T_C$ suppression depends strongly on gap anisotropy. It also contrasts with $T_C$ suppression in $Ba_{1-x}K_xFe_2As_2$ samples with differing impurity levels, suggesting that columns generate predominantly intraband scattering. For both columnar defects and point impurities[28], the penetration depth changes from exponential to $\Delta\lambda \sim T^2$ as the defect density is raised. These findings indicate that pure $Ba_{0.6}K_{0.4}Fe_2As_2$ pairs in an $s_{+-}$ state that evolves, under increasing defect density, to a configuration with midgap states.

**Acknowledgements**

We wish to thank thank P. Hirschfeld for emphasizing the role of gap anisotropy in reducing $T_C$. We also thank A. Chubukov, H. Kontani and A.V. Boris for useful discussions. Work at Argonne and UIUC was supported by the Center for Emergent Superconductivity, an Energy Frontier Research Center funded by the US Department of Energy, Office of Science, Office of Basic Energy Sciences under Award No. DE-AC0298CH1088. Work at The Ames Laboratory was supported by the U.S. Department of Energy, Office of Basic Energy Sciences, Division of Materials Sciences and Engineering under contract No. DE-AC02-07CH11358. A.K. Ghosh was supported in part by the Indo-US Science and Technology Forum, Work in China was supported by the MOST of China (2011CBA00102, 2012CB821403) and PAPD.

Corresponding author: russg@illinois.edu

[1] I.I. Mazin, D.J. Singh, M.D. Johannes, M.H. Du, Phys. Rev. Lett. **101**, 057003 (2008)

[2] K. Kuroki et. al., Phys. Rev. Lett. **101**, 087004 (2008)

[3] R. Prozorov, V. G. Kogan, Rep. Prog. Phys. **74**, 124505 (2011).

[4] Y. Bang, Euro. Phys. Lett. **86**, 47001 (2009)

[5] A. B. Vorontsov, M.G. Vavilov, A.V. Chubukov, Phys. Rev. B **79**, 140507 (2009)

[6] S. Onari and H. Kontani, Phys. Rev. Lett. **103**, 177001 (2009).

[7] D.V. Efremov, M.M. Korshunov, O.V. Dolgov, A.A. Golubov, P.J. Hirschfeld, Phys. Rev. B **84**, R180512 (2011)

[8] Y. Wang, A. Kreisel, P. J. Hirschfeld and V. Mishra, cond-mat arXiv:1210.7474 (2012).

[9] K. Hashimoto, A. Serafin, S. Tonegawa, R. Katsumata, R. Okazaki, T. Saito, H. Fukazawa, Y. Kohori, K. Kihou, C. H. Lee, A. Iyo, H. Eisaki, H. Ikeda, Y. Matsuda, A. Carrington, and T. Shibauchi, Phys. Rev. B **82**, 014526 (2010)

[10] J.-Ph. Reid, M. A. Tanatar, A. Juneau-Fecteau, R. T. Gordon, S. René de Cotret, N. Doiron-Leyraud, T. Saito, H. Fukazawa, Y. Kohori, K. Kihou, C. H. Lee, A. Iyo, H. Eisaki, R. Prozorov, L. Taillefer, Phys. Rev. Lett. **109**, 087001 (2012)

[11] Y. Nakajima, Y. Tsuchiya, T. Taen, T. Tamegai, S. Okayasu, Phys. Rev. B **80**, 012510 (2009)

[12] L. Fang, Y. Jia, C. Chaparro, G. Sheet, H. Claus, M. A. Kirk, A. E. Koshelev, U. Welp, G. W. Crabtree, W. K. Kwok, S. Zhu, H. F. Hu, J. M. Zuo, H.-H. Wen, B. Shen, Appl. Phys. Lett. **101**, 012601 (2012).

[13] M. Rotter, M. Tegel, D. Johrendt, Phys. Rev. Lett. **101**, 107006 (2008)

[14] H. Kim, R. T. Gordon, M. A. Tanatar, J. Hua, U. Welp, W. K. Kwok, N. Ni, S. L. Bud'ko, P. C. Canfield, A. B. Vorontsov, and R. Prozorov, Phys. Rev. B **82**, 060518 (2010).

[15] L. Shan, Y.-L. Wang, J. Gong, B. Shen, Y. Huang, H. Yang, C. Ren, and H.-H. Wen, Phys. Rev. B **83**, R060510 (2011)

[16] A. Carrington, R.W. Giannetta, J.T. Kim, J. Giapintzakis, Phys. Rev. B **59**, R14173 (1999)

[17] R. Prozorov, R. W. Giannetta, A. Carrington, and F. M. Araujo-Moreira, Phys. Rev. B **62**, 115 (2000).

[18] R. Prozorov and R.W. Giannetta, Supercond. Sci. Technol. **19**, R41 (2006)

[19] A. Charnukha, O.V. Dolgov, A.A. Golubov, Y. Matiks, D.L. Sun, C.T. Lin, B. Keimer, A.V. Boris, Phys. Rev. B **84**, 174511 (2011)

[20] M. A. Tanatar, J. P. Reid, H. Shakeripour, X. G. Luo, N. Doiron-Leyraud, N. Ni, S. L. Bud'ko, P. C. Canfield, R. Prozorov, and L. Taillefer, Phys. Rev. Lett. **104**, 067002 (2010)

[21] J. P. Reid, M. A. Tanatar, X. G. Luo, H. Shakeripour, N. Doiron-Leyraud, N. Ni, S. L. Bud'ko, P. C. Canfield, R. Prozorov, and L. Taillefer, Phys. Rev. B **82**, 064501 (2010).

[22] C. Martin, H. Kim, R. T. Gordon, N. Ni, V. G. Kogan, S. L. Bud'ko, P. C. Canfield, M. A. Tanatar, and R. Prozorov, Phys. Rev. B **81**, 060505 (2010).

[23] A.A. Abrikosov and L.P. Gor'kov, Zh. Eksp. Teor. Fiz. **39**, 1781 (1960); Sov. Phys. JETP **12**, 1243 (1961)

[24] V.G. Kogan, Phys. Rev. B **80**, 214532 (2009)

[25] A. A. Golubov and I. I. Mazin, Phys. Rev. B **55**, 15146 (1997)

[26] L. A. Openov, JETP Lett. **66**, 661 (1997)

[27] We wish to thank P.J. Hirschfeld for suggesting this scenario.

[28] K. Hashimoto, T. Shibauchi, S. Kasahara, K. Ikada, S. Tonegawa, T. Kato, R. Okazaki, C. J. van der Beek, M. Konczykowski, H. Takeya, K. Hirata, T. Terashima and Y. Matsuda, Phys. Rev. Lett. **102**, 207001 (2009).

[29] R.T. Gordon, H. Kim, N. Salovich, R. Giannetta, R.M. Fernandes, V.G. Kogan, T. Prozorov, S.L. Bud'ko, P.C. Canfield, M.A. Tanatar, R. Prozorov, Phys. Rev. B **82**, 054507 (2010)

[30] H. Chen, Y. Ren, Y. Qiu, Wei Bao, R. H. Liu, G. Wu, T. Wu, Y. L. Xie, X. F. Wang, Q. Huang, X. H. Chen, Europhys. Lett. **85**, 17006 (2009)

[31] H. Fukazawa, T. Yamazaki, K. Kondo, Y. Kohori, N. Takeshita, P. M. Shirage, K. Kihoui, K. Miyazawa, H. Kito, H. Eisaki, A. Iyo , Jour. Phys. Soc. Jpn. **78**, 033704 (2009)

[32] R.M. Fernandez, M.G. Vavilov and A.V. Chubukov, Phys. Rev. B **85**, R140512 (2012)

[33] R.M. Fernandez, J. Schmalian, Phys. Rev. B **82**, 140521 (2010)

[34] J.D. Fletcher, A. Carrington, O.J. Taylor, S.M. Kasakov, J. Karpinski, Phys. Rev. Lett. **95**, 097005 (2005)

[35] N. Chikumoto, A. Yamamoto, M. Konczykowski, M. Murakami, Physica C **378** 381 (2002)

[36] I. I. Mazin, O. K. Andersen, O. Jepsen, O.V. Dolgov, J. Kortus, A. A. Golubov,A. B. Kuz'menko,and D. van der Marel, Phys. Rev. Lett. **89**, 107002 (2002)

[37] I. I. Mazin and V.P. Antropov, Physica C **385**, 49 (2003)